# Novel Diamond Anvil Cell for Electrical Measurements using Boron-doped Metallic Diamond Electrodes

R. Matsumoto,[1,2] Y. Sasama,[1,2] M. Fujioka,[1,3] T. Irifune,[4] M. Tanaka,[1] T. Yamaguchi,[1,2] H. Takeya,[1] and Y. Takano[1,2]

[1] *MANA, National Institute for Materials Science, Tsukuba 305-0047, Japan*
[2] *Graduate School of Pure and Applied Sciences, University of Tsukuba, Tsukuba 305-8577, Japan*
[3] *Laboratory of Nano-Structure Physics, Research Institute for Electronic Science, Hokkaido University, Sapporo 001-0020, Japan*
[4] *Geodynamics Research Center, Ehime University, Matsuyama 790-8577, Japan*

A novel diamond anvil cell suitable for electrical transport measurements under high pressure has been developed. A boron-doped metallic diamond film was deposited as an electrode onto a nano-polycrystalline diamond anvil using a microwave plasma-assisted chemical vapor deposition technique combined with electron beam lithography. The electrical transport measurements of Pb were performed up to 8 GPa, and the maximum pressure reached was above 30 GPa. The boron-doped metallic diamond electrodes showed no signs of degradation after repeated compression measurements.

## 1. Introduction

High-pressure techniques have been developed in the field of geophysics to investigate the inner structure of the earth [1,2]. Physical property measurements under high pressure have become quite important in a number of different research fields, such as materials science [3]. In particular the superconducting properties of many materials under high pressure have recently attracted considerable attention. A record of superconducting transition temperature $T_c$ at 203 K was recently established under 150 GPa in hydrogen sulfide [4]. It is theoretically predicted that hydrogen will become metallic under extremely high pressures of ~400 GPa and exhibits superconductivity at room temperature [5].

The discovery of superconductivity in hydrogen sulfide was accomplished under high pressure using a diamond anvil cell (DAC) [6,7] which is the most widely used device for high pressure generation at present. To evaluate the superconducting properties under high pressure, electrical transport measurements using the DAC are necessary. However, it is difficult because of the necessity of a small sample sizes (< 100 μm) and the deformation of electrodes under compression [8]. The development of an innovative technique to easily perform electrical transport measurements under high pressure is required.

In this study, we have developed a novel DAC for electrical transport measurements under high pressure. Key components of the DAC are a boron-doped metallic diamond electrode [9,10] and a nano-polycrystalline diamond (NPD) anvil [11]. The heavily boron-doped diamond thin film can be synthesized by a microwave plasma-assisted chemical vapor deposition (MPCVD) method [12]. When the boron concentration exceeds $3 \times 10^{20}$ $cm^{-3}$, diamond shows metallicity and superconductivity at low temperature [12-14]. The NPD consists of nanoparticle, leading to the high mechanical hardness [15]. The DAC can perform electrical transport measurements easily using microscale boron-doped diamond electrode fabricated on the NPD anvil. Furthermore, this metallic diamond electrode on the NPD can be used repeatedly until the anvil itself broken.

## 2. Fabrication of electrodes on anvil

Fig. 1 (a) illustrates the schematic of a designed DAC, comparing with a conventional DAC (b). The DAC is composed of a square NPD anvil with a metallic diamond electrode and a general culet diamond. The fabrication of the electrode was performed using the following steps [16,17]. (1) The shape of electrode was designed by a resist using electron beam lithography in a combination with a scanning electron microscope (JEOL: JSM-5310) associated with a nanofabrication system (Tokyo Technology: Beam Draw). (2) Ti/Pt metal mask was deposited on the designed resist. (3) The excess region of the metal mask was removed using a lift-off process. (4)

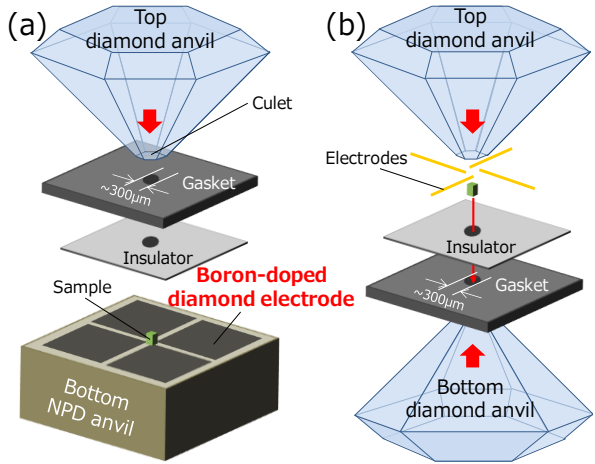

Fig. 1. (a) Novel DAC assembly with metallic boron-doped diamond electrode for the electrical transport measurement. (b) Conventional DAC assembly.

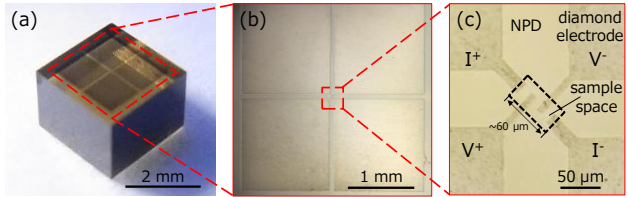

Fig. 2. (a) Optical image of the developed diamond anvil. (b) Top view of the diamond electrode. (c) Enlargement of the four-terminal area.

The boron-doped diamond was selectively deposited on the uncovered region of the NPD surface. After 2.5 min of the deposition, an electrode of 100 nm thickness was obtained. The detailed conditions of the deposition are presented in reference [9].

The four-terminal of the diamond electrode for resistivity measurements was designed on the NPD anvil. The optical image of the electrode is shown in Fig. 2 (a). The top view of the electrode and enlargement of four-terminal at the sample space are shown in Fig. 2 (b,c).

## 3. Transport measurements under high pressure

We show electrical transport measurements under high pressure using designed anvil described in Fig. 2. The resistance of Pb was measured by a standard four-terminal method using a physical property measurement system (Quantum Design: PPMS). The sample was deposited on the four-terminal of electrode in the sample space. The culet size of the top anvil was 0.6 μm. A stainless steel sheet with a thickness of 200 μm was used as a gasket. The gasket was electrically insulated from the electrode by cubic boron nitride (cBN) powder. The ruby powder is mixed in a pressure-transmitting medium of cBN powder. The pressure values were determined from the peak position of the ruby fluorescence [18]. The fluorescence was detected by raman spectroscopy system (RENISHAW: inVia Raman Microscope).

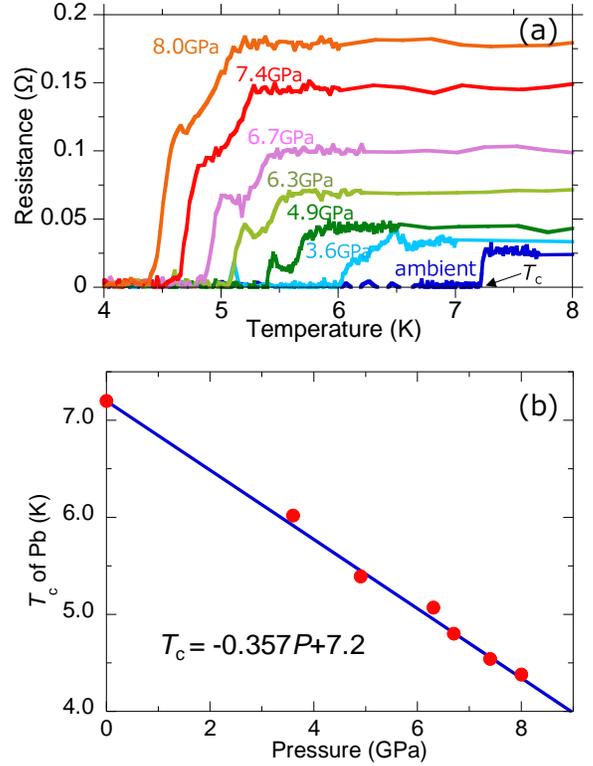

Fig. 3. (a) Temperature dependence of resistivity for Pb under several pressures. (b) Pressure dependence of $T_c$ for Pb.

## 4. Results and discussion

Fig. 3 (a) shows the temperature dependence of the electrical resistance of Pb under several pressures. The resistivity exhibits metallic behavior, and suddenly dropped to zero which corresponds to the superconducting transition of around 7.2 K at ambient pressure. The temperature at zero resistance is gradually decreased from 7.2 K to 4.4 K with increase of applied pressure up to 8 GPa. The pressure dependence of $T_c$ is plotted in Fig. 3 (b), and $dT_c/dP$ is estimated to be -0.357 K/GPa using a least-squares method. This value shows good agreement with previous report [19].

To generate higher pressures up to 30 GPa, we chose a top anvil with a culet size of 0.4 mm. Fig. 4 shows ruby fluorescence spectra at various stroke lengths. The zero point of the stroke length was defined as a contact position of the culet and the gasket. The fluorescence associated with the peaks of $R_1$ and $R_2$ around 700 nm [6], and the peak position of $R_1$ was used to determine the pressure value. The peak position of $R_1$ moved to higher wavelength from 694.3 nm with increase of the stroke length, and reached 705.3 nm at a stroke length of 230 μm. The pressure value $P$ was estimated from wavelength shift $\Delta\lambda$ by following relation: $P = \Delta\lambda / 0.365$ GPa [20]. The obtained pressure values were plotted in Fig. 5 as a function of the stroke length. The maximum pressure reached was above 30 GPa. Considering the steep curve

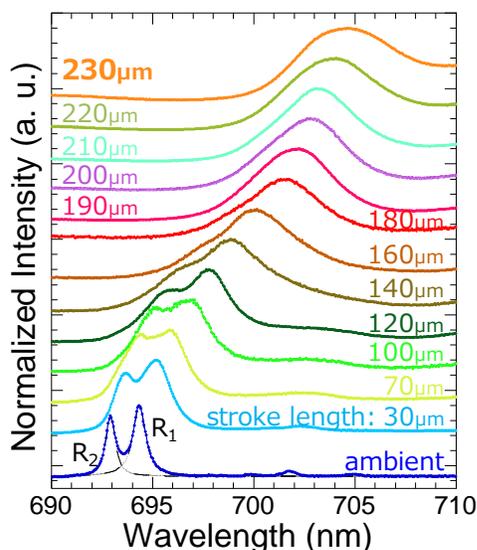

Fig. 4. Peak shift of ruby fluorescence by increase in stroke length.

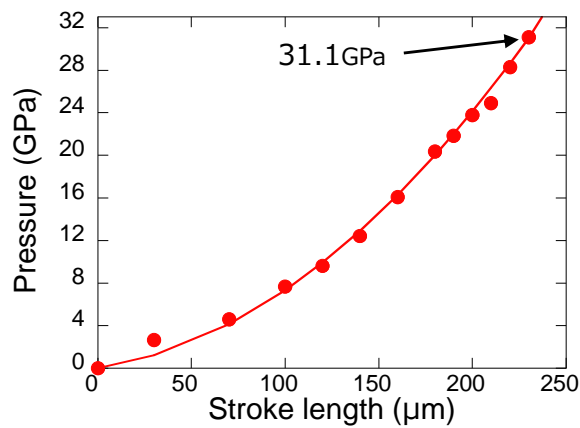

Fig. 5. Estimated pressure by ruby fluorescence at various stroke lengths.

around 30 GPa, the DAC will be able to reach higher pressure.

## 5. Summary

The novel DAC for electrical transport measurements was successfully developed by using the boron-doped metallic diamond electrode on the NPD anvil. The electrode was fabricated by combination of the electron beam lithography technique and the MPCVD method. The electrical transport measurement for Pb was performed up to 8 GPa. The maximum pressure of the DAC reached above 30 GPa at several times without any degradation in the metallic diamond electrodes.

**Acknowledgements**

The authors thank Dr. T. Sakai (Ehime University) and Dr. S. J. Denholme (Tokyo University of Science) for helpful discussions and proofreading of a manuscript.

A part of the fabrication process was supported by NIMS Nanofabrication Platform in Nanotechnology Platform Project sponsored by the Ministry of Education, Culture, Sports, Science and Technology (MEXT), Japan. The preparation of NPD substrate was supported by the Visiting Researcher's Program of Geodynamics Research Center, Ehime University.